\documentclass[12pt,a4paper]{article}
\newlength{\subfigcolsep}
\makeatletter 
 
\@addtoreset{figure}{section} 
\makeatother 

\usepackage{lscape}
\usepackage[dvipdfm]{graphicx}
\usepackage{amsmath}
\usepackage{amssymb}
\usepackage{bm}
\usepackage{times}
\usepackage{multicol}
\usepackage{setspace}
\usepackage{subfigure}
\usepackage{url}
\newtheorem{ theorem name}{ caption}
\newtheorem{ proof name}{ caption}

\makeatletter

\@addtoreset {equation} {section}
\makeatother

\begin{document}
\begin{center}
\LARGE{Stability and Restoration phenomena in Competitive Systems}
\end{center}
\vspace{1.0cm}

\centerline{Lisa Uechi\footnote{E-mail: uechir@kuicr.kyoto-u.ac.jp} and
Tatsuya Akutsu\footnote{E-mail: takutsu@kuicr.kyoto-u.ac.jp}}

\vspace{0.5cm}

\centerline{$^{1,2}$ Bioinformatics Center, Institute for Chemical
Research,}
\centerline{
Kyoto University, Gokasho, Uji, Kyoto 611-0011, Japan}

\vspace{1.5cm}

\setcounter{page}{0}
\thispagestyle{empty}

\setcounter{page}{0}
\thispagestyle{empty}
\pagenumbering{arabic}

\begin{center}
{\bf Abstract}\\
\end{center}
A conservation law and stability, recovering phenomena and characteristic patterns of a nonlinear dynamical system have been studied and applied to biological and ecological systems. In our previous study, we proposed a system of symmetric $2n$-dimensional conserved nonlinear differential  equations with external perturbations. In this paper, competitive systems described by $2$-dimensional nonlinear dynamical (ND) model with external perturbations are applied to population cycles and recovering phenomena of systems from microbes to mammals.

The famous 10-year cycle of population density of Canadian lynx and snowshoe hare is numerically analyzed. We find that a nonlinear dynamical system with a conservation law is stable and generates a characteristic rhythm (cycle) of population density, which we call the {\it standard rhythm} of a nonlinear dynamical system. The stability and restoration phenomena are strongly related to a conservation law and balance of a system. The {\it standard rhythm} of population density is a manifestation of the survival of the fittest to the balance of a nonlinear dynamical system.

\section{Introduction}
The concept of stability is important in order to understand natural
phenomena in biological and engineering systems. In our previous study, we studied the relation between a conservation law and stability of a $2n$-dimensional competitive system that contains competitive interactions, self-interactions and mixing interactions. The system consists of $2n$-dimensional nonlinear differential equations required from
Noether's theorem \cite{uechi2012conservation}.
The $2n$-dimensional nonlinear ordinary differential equations for a competitive
system constructed to satisfy the conservation law have properties such as 
the addition law, which is empirically interpreted as recovery from injuries of skin and tissues in biological bodies.

It has been shown by many researchers that a relatively simple set of interactions can explain complex phenomena in biological systems \cite{meinhardt1982models,gierer1972theory}.
For example, in 1952, Turing suggested chemical molecular mechanism called the reaction-diffusion system \cite{turing1952chemical} which is defined as semi-linear parabolic partial differential equations. This reaction-diffusion system is well applied for explaining stripe patterns of the marine angelfish, {\it Pomacanthus}, and restoration phenomena in its stripe patterns from injuries was observed \cite{kondo1995reaction,lengyel1991modeling,kumar2011effects}. Prigogine also proposed Brusselator model with nonlinear ordinary differential equations to illustrate spatial oscillations and Turing patterns \cite{PhysRevE.84.026201}. It is also an interesting problem to investigate in ecological systems if a large complex system should be stable or not, and many researchers have discussed the criteria concerning the stability of a system for $n$ dimensional ordinary differential equations and statistical framework \cite{may1972will,cohen1985will,tokita2004species,daniels1974stability,ives2007stability,allesina2012stability,m2012sexual}.
 What would be a reason why a simple set of interactions can explain complex phenomena?
We discussed a system of interactions generalizing Lotka-Volterra type
nonlinear competitive interactions and suggested that a conservation law could be a key to
understand complex phenomena even in biological and ecological systems.

We investigated the system of $2n$-dimensional coupled first-order
differential equations by using Noether's theorem, which led to the following results.
(i) The form of differential equations and coefficients of nonlinear interactions are strictly confined when the system has a conservation law which is constructed by interacting species of a particular experimental system. (ii) The conserved
quantity of a system produces a Lyapunov function which is usually
employed to study solutions of nonlinear differential equations. 
The conserved quantity is constructed by Noether's theorem, but the analysis of Lyapunov function would be used to check solutions to differential equations including those for non-conservative and dissipative systems. The system of differential equations with conservation law is different in this respect.
(iii) A system of interactions could be analyzed as an assembly of a basic binary-coupled form (BCF).
In other words, a complex interacting system can be decomposed into an
assembly of binary-coupled systems. The BCF system is a simple basic set to explain complex phenomena defined by Noether's theorem.  (iv) The BCF system with conservation law indicates an addition law which may be interpreted as
the restoration or rehabilitation phenomena; those are known in a large system of
neural network or computer network when a small
disordered device or a part of network system is replaced by a
normal device. These properties could be applied to stability and restoration phenomena of biological systems.
(v) The conservation law is also useful to check accuracy of
numerical solutions to nonlinear differential equations. 
As summarized above, we discussed that the basic nonlinear system in BCF is stable. The binary-coupled system and addition law supported by a conservation law can lead to a large, stable complex system. This is an important conclusion in the conserved binary-coupled model. Because the BCF system has such several interesting properties, we will apply the model in order to study stability and interaction mechanism of a biological system.

In this paper, we will explain the properties of solutions with a conservation law and applications to biological systems. In Section \ref{Computational Simulation}, we extend the BCF model to simulate external perturbations numerically. There are various prey-predator type competitive models with perturbations \cite{khaminskii2003some,martinez2011predator,shiang2009perturbation,chen2001transient,khasminskii2001long,rudnicki2007influence,froda2007prediction},  however, most of them are with small, stochastic perturbations. The behaviors of conservation laws with external perturbations have been seldom considered. We will explicitly discuss properties of the conserved, stable, 2-variable nonlinear interacting system with external perturbations and the conservation law, its indications and possible applications to nonlinear interacting system. We will show that  2-variable ND model has the properties of restoration and recovery from external perturbations. 
In Section \ref{Conservation law with external perturbation}, stability and population cycles of biological systems are examined in terms of a conservation law of the system. We will also examine specific examples of the Canadian lynx and snowshoe hare \cite{stenseth1998patterns,elton1942ten,stenseth1997population,krebs2001drives,basille2011ecologically,blasius1999complex,maquet2007global} and the question of population cycles, and food chain of microbes in the lake \cite{schindler2012mysis,gazi2012dynamics}.
Conclusions and summary of results are given in Section \ref{Conclusion and Discussion}.

\section{The model of binary-coupled form (BCF)}\label{Computational Simulation}
\subsection{$2n$-ND system with perturbations}\label{2n-ND system with perturbations}
We discussed BCF system and the conservation law of $2n$-nonlinear dynamical ($2n$-ND) model in detail in the previous work \cite{uechi2012conservation}.
In this study, we include external perturbations in $2n$-variable nonlinear differential equations in order to examine characteristic behaviors of conserved nonlinear interacting systems.
It should be noticed that the $2n$-ND model is extended by adding external perturbation terms which maintain a conservation law given by Noether's theorem. The odd variable terms for $x_{i}$ ($i=1,\dots,2n$) are
\begin{eqnarray}
\begin{split}
d_{2k,2k-1}\dot{x}_{2k-1} =& \sum_{i=1}^n \Bigl\{  ( \alpha_{\{ 2ni +2k \}} + \alpha_{ \{2n^2  
+2nk + 2i-1 \}} ) x_{2i-1} + (\alpha_{\{2n^2 +2ni +2k\}} + \alpha_{\{2n^2  +2nk +2i \}})x_{2i}
\\&+\alpha_{\{ 4n^2 +2ni +2k \}}x_{2i-1}x_{2i} \Bigr\}
+\sum_{j=1}^{2n} \alpha_{\{4n^2 +2nk + j\}}x_jx_{2k-1}+c_{2k-1} ,\label{2n_ND_2k-1}
\end{split}
\end{eqnarray}
where $k=1,\dots,n$.
The even variable terms for $x_{i}$ ($i=1,\dots,2n$) are
\begin{eqnarray}
\begin{split}
d_{2k-1,2k} \dot{x}_{2k} =& \sum_{i=1}^n \Bigl\{ (\alpha_{\{2ni+2k-1\}} + \alpha_{\{2nk
+2i-1\}}) x_{2i-1} +  (\alpha_{\{2n^2 + 2ni +2k-1\}} + \alpha_{\{2nk +2i\}})x_{2i}\\
& +  \alpha_{\{4n^2 +2ni +2k-1\}} x_{2i-1} x_{2i} \Bigr\}+ \sum_{j=1}^{2n} \alpha_{\{4n^2 +2nk+j\}}x_j x_{2k}  + c_{2k},\label{2n_ND_2k}
\end{split}
\end{eqnarray}
where $\dot{x} = dx/dt$, coefficients, $d_{i,j}$ express $d_{2k,2k-1}=\alpha_{2k}-\alpha_{2k-1}$, $d_{2k-1,2k}=\alpha_{2k-1} - \alpha_{2k}$. The linear coefficients and nonlinear coefficients $\alpha_i$, ($i=1,\dots,8n^2 +2n$) are arbitrary constant values.
The last terms $c_{2k-1}$, $c_{2k}$, $(k=1,\dots,n)$ of \eqref{2n_ND_2k-1} and \eqref{2n_ND_2k} are constant or piecewise continuous constant, which are interpreted as external perturbations (temperature, seasons and other temporal, external inputs). One should note that constant terms have dimension of velocity, so they are different from actual external perturbations which are considered to effectively express external perturbations. Because external perturbations (inputs) change population densities as $\dot{x}=dx/dt$, we simulate numerically those effects with $c_{2k-1}$, $c_{2k}$ as external inputs. The system has a conservation law derived from Noether's theorem which is proved in the paper \cite{uechi2012conservation}: 
\begin{eqnarray}
\begin{split}
\Psi \equiv  \sum_{i=1}^n \sum_{j=1}^{2n} \Bigl\{ \alpha_{\{ 2ni +j\}}x_{2i-1} x_j 
+ \alpha_{\{ 2n^2  +2ni +j \}} x_{2i}x_j + \alpha_{ \{4n^2 +2ni +j \}} x_{2i-1}x_{2i}x_j \Bigr\}
\\ +\sum_{i=1}^n \{ c_{2i} x_{2i-1} + c_{2i-1} x_{2i} \}.\label{cons_2n_ND}
\end{split}
\end{eqnarray}
Therefore, with the equations from \eqref{2n_ND_2k-1} to \eqref{cons_2n_ND}, we are able to consider the conserved nonlinear dynamical system with external perturbations by employing piecewise continuous constant terms, $c_{2k-1}$, $c_{2k}$.
\subsection{Properties of 2-variable ND model}\label{Recovering of 2-variable ND model}
The equations of 2-variable ND model are produced by setting $n=1$ ($k=1$) in equations \eqref{2n_ND_2k-1}  to \eqref{cons_2n_ND}, resulting in
\begin{eqnarray}
\dot{x}_1 = \frac{1}{d_{21}}\{ (\alpha_4 + \alpha_5)x_1 + 2\alpha_6 x_2 
+ 2\alpha_8x_1x_2 + \alpha_7x_1^2 \} + \frac{c_1}{d_{21}},\label{2_var_x1}
\end{eqnarray}
\begin{eqnarray}
\dot{x}_2 = \frac{1}{d_{12}} \{ 2\alpha_3 x_1 + (\alpha_4 + \alpha_5) x_2 +
2\alpha_7 x_1 x_2 + \alpha_8x_2^2 \} + \frac{c_2}{d_{12}},\label{2_var_x2}
\end{eqnarray}
and the 2-variable ND model has the following conservation law,
\begin{eqnarray}
\Psi \equiv \alpha_3 x_1^2 + (\alpha_4 + \alpha_5)x_1x_2 + \alpha_6x_2^2 
+ \alpha_7 x_1^2 x_2 + \alpha_8x_1 x_2^2 +c_2x_1 + c_1 x_2.\label{2_var_cons}
\end{eqnarray}

\begin{table}[b!]
\centerline{
Table 1: The list of nonlinear coefficients.}\vspace{0.2cm}
\begin{center}
  \begin{tabular}{c|c |c| c| c | c|  c| c| c} \hline
&  $\alpha_1$  & $\alpha_2$ & $\alpha_3$ & $\alpha_4$ & $\alpha_5$ &$\alpha_6$ & $\alpha_7$ & $\alpha_8$ \ \\  \hline \hline
Condition 1&1.0 & 401.0  & -0.35 & 1.5 & 10.5 & -0.01 & -0.006 & -0.011 \\ \hline
Condition 2&1.0 & 401.0  & -0.35 & 1.5 & 10.5 & -0.51 & -0.006 & -0.011 \\ \hline 
     \end{tabular}
\end{center}\label{table1}
\end{table}
The nonlinear interactions can generally represent, for example, Lotka-Volterra type prey-predator, competitive interactions, food-chain relations by adjusting nonlinear parameters $\alpha_1, \dots, \alpha_8$. The piecewise continuous constants, $c_1$ and $c_2$ are used as external perturbations in computer simulations, such as environmental conditions, food or hormones which increase or decrease interacting species in questions. The equations \eqref{2_var_x1} $\sim$ \eqref{2_var_cons} form 2-variable BCF nonlinear differential equations with a conservation law.

By employing eqs. \eqref{2_var_x1} $\sim$ \eqref{2_var_cons}, we will show: \\
(1) solutions to the binary-coupled nonlinear equations maintain a characteristic $(x_1,x_2)$ phase-space of solutions and recovery from external perturbations. The external perturbations can numerically reproduce environmental conditions such as temperature and climate, environmental hormones and chemical substances which affect interacting species. The nonlinear binary-coupled model can be applied to examine responses of a system whether they are induced from internal interactions or external perturbations.\\
(2) The binary-coupled nonlinear equations with conservation law exhibit stable phase-space solutions, which are interpreted as stability and recovery of population-change in a biological system. The properties of the binary-coupled nonlinear interactions will be shown explicitly in numerical simulations.\\
(3) By employing the 2-variable binary-coupled model, it is possible to simulate cycles of maxima and minima in population-change, delays of periodic times of population cycles for competitive species \cite{khaminskii2003some}. Hence, cycles of population-change will be discussed in terms of the conservation law and nonlinear interactions.

\begin{figure}[h]
\begin{center}
 \subfigure[2-variable ND solutions. Solid and dashed lines represent $x_1$ (prey) and $x_2$ (predator), respectively. One should note that the unit of time should be defined with respect to a system in consideration.]{%
   \includegraphics[width=.45\columnwidth]{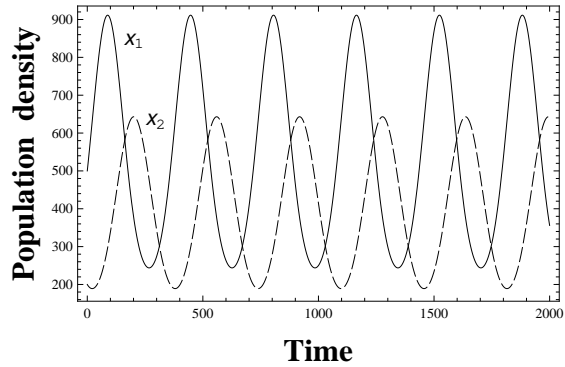}\label{2_ND_sim_one_pert_sol_ref}} \\
\end{center}
 \subfigure[Phase-space of 2-variable ND solutions.]{%
  \includegraphics[width=.43\columnwidth]{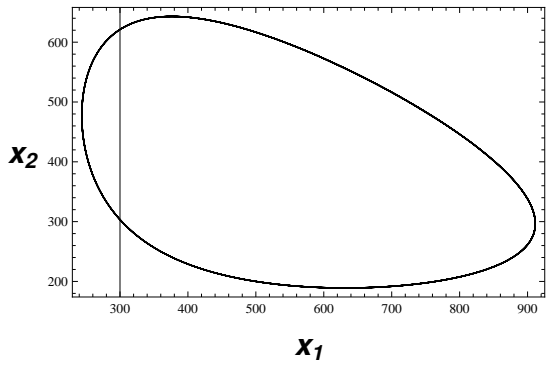}\label{2_ND_sim_one_pert_two_sol_ref}} \hspace{10mm}
 \subfigure[Conservation law $\Psi$ of 2-variable ND. It is constant with respect to time.]{%
   \includegraphics[width=.45\columnwidth]{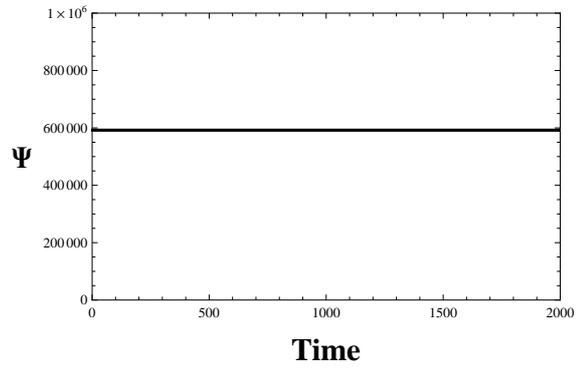}\label{2_ND_sim_one_pert_con_ref}}
\caption{A 2-variable ND solution and Conservation law $\Psi$.}\label{2_ND_ref}
\end{figure}

Figure \ref{2_ND_sim_one_pert_sol_ref} shows the nonlinear interactions between species without external perturbations ($c_1=0$ and $c_2=0$), whose coefficients of nonlinear equations are set as in Table 1 (Condition 1). In a view of the classical Lotka-Volterra competitive system, it can be interpreted as that $x_1$ and $x_2$ represent prey and predator, respectively. Figure \ref{2_ND_sim_one_pert_two_sol_ref} is the phase-space given by solutions $(x_1, x_2)$. The solutions $(x_1, x_2)$ are periodic with respect to time, the maximum and minimum of $(x_1, x_2)$ appear with a time-delay. Figure \ref{2_ND_sim_one_pert_con_ref} shows the numerical value of the conserved function $\Psi$ defined by \eqref{2_var_cons}, which is constant with respect to time.

The solutions $(x_1, x_2)$ in Figure \ref{2_ND_sim_one_pert_sol_ref} show explicitly a time-delay of the peak for interacting species. The timings of peak and delayed peak are determined by nonlinear interactions and strength of coupling constant. The solutions $(x_1, x_2)$ in Figure \ref{2_ND_sim_one_pert_two_sol_ref} show phase-space solutions, which are stable in the meaning that the conserved quantity $\Psi$ is maintained constant and phase-space solutions are in the same trajectory for all time. The unit of time should be considered to adjust to the time scale of a system in consideration, because biological unit times are generally different from microbes to mammals.

The phase-space diagram \ref{2_ND_sim_one_pert_two_sol_ref} and the straight line of Figure \ref{2_ND_sim_one_pert_con_ref} show that the solution is exact and stable \cite{uechi2012conservation}. The three figures exhibit important properties of solutions to the system of prey-predator type of competitive nonlinear interactions.

One of the important properties shown by the stable, conserved nonlinear system is that the interacting species repeat the rhythm of maxima and minima of the population. The periods of the rhythm are the result of complicated nonlinear interactions, but the system keeps the constant quantity $\Psi$ with respect to time. The interesting applications of the BCF model are shown by employing in the paper \lq {\it Mysis} in the Okanagan Lake food web \rq \cite{schindler2012mysis}, Canadian Lynx and snowshow hare \cite{elton1942ten}, which will be explained in Section 3.

\subsection{Recovering and restoration from perturbations}\label{Recovering and restoration from perturbations}
In order to investigate responses of a system to external perturbations, we introduce piecewise continuous constants, $c_1$ and $c_2$, by using $\theta$-functions such that
\begin{eqnarray}
\begin{split}
c_i = f_i \{ \theta(t-t_{start}) - \theta(t-t_{end}) \},\ \ \ (i=1,2),\label{constant}
\end{split}
\end{eqnarray}
where $\theta(t-t')$ represents a step function:
\begin{equation}
 \theta(t-t') = \begin{cases}
              1, &    (t \ge t'),  \\
              0, &    (t < t'),
              \end{cases}\label{theta}
\end{equation}
and coefficients $f_i , (i=1,2)$ are positive or negative constants to express strength of external perturbations. The constants are adjusted to produce reasonable maxima and minima in numerical simulations.

\begin{figure}[h]
\begin{center}
 \subfigure[2-variable ND solutions with a negative perturbation on prey $x_1$. The perturbation is introduced from $t=700$ to $t=1200$ which is represented as gray background.]{%
   \includegraphics[width=.45\columnwidth]{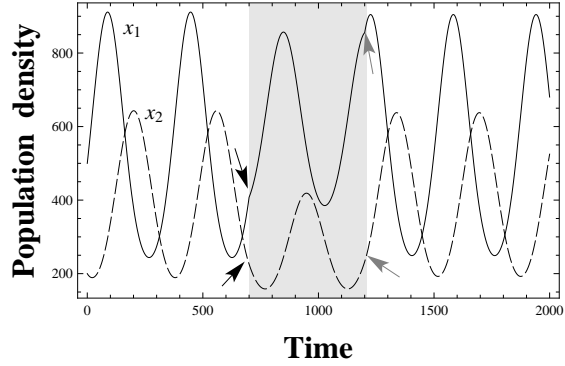}\label{2_ND_sim_prey_onep_same_sol}}\\
\end{center}
 \subfigure[The ($x_1$,$x_2$) phase-space transition with the negative perturbation as in (a). Solid line (St.~1) is initial state, and St. ~2 is the recovered state after the end of perturbation. Dashed line is phase-space during Sp.~1 - Ep.~1.]{%
	\includegraphics[width=.43\columnwidth]		{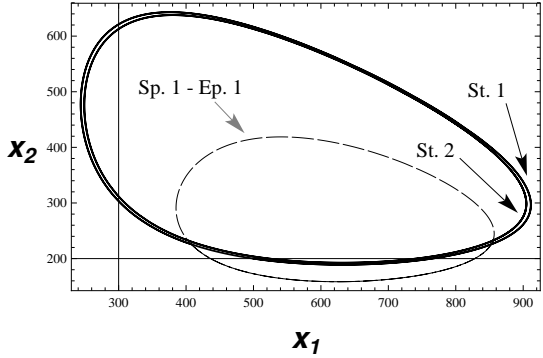}\label{2_ND_sim_prey_onep_same_two_sol}}
\hspace{10mm}
 \subfigure[Conservation law $\Psi$ of 2-variable ND with one perturbation. $\Psi$ changed $\Psi \simeq 60000$ to $\Psi \simeq 30000$ by introducing perturbation. It recovers after Ep. 1.]{%
	\includegraphics[width=.45\columnwidth]{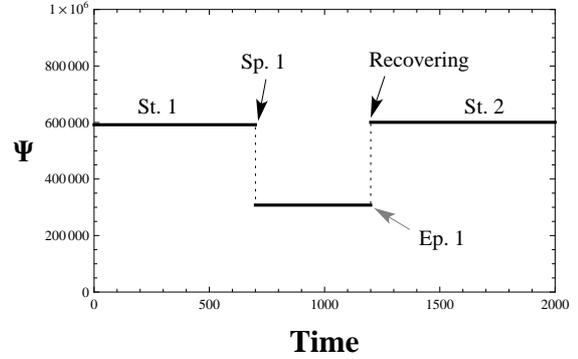}\label{2_ND_sim_prey_onep_same_cons}}
\caption{An external perturbation and a recovery.}\label{2_var_nd}
\end{figure}

\begin{figure}[h]
 \subfigure[2-variable ND solutions with a critical negative perturbation on prey, $x_1$. Solutions converge to zero after the perturbation.]{%
   \includegraphics[width=.43\columnwidth]{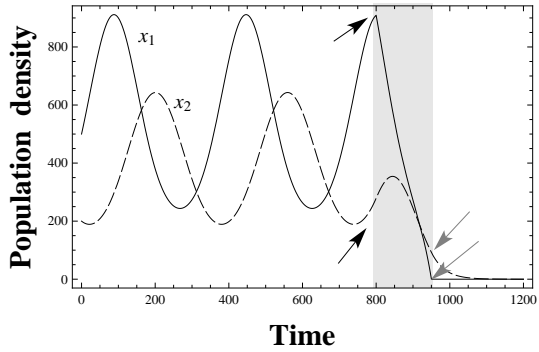}\label{2_ND_sim_onep_ex_sol}}\hspace{10mm}
 \subfigure[Conservation law $\Psi$ with a critical negative perturbation. $\Psi$ converges to zero after the critical perturbation.]{%
   \includegraphics[width=.45\columnwidth]{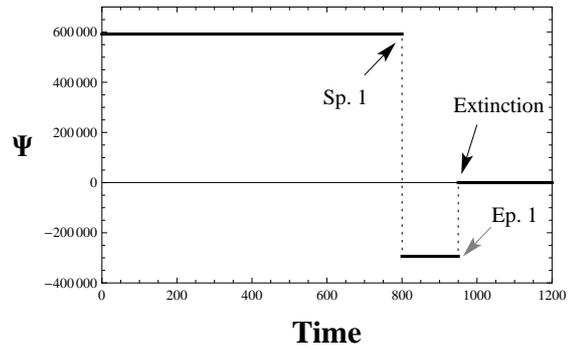}\label{2_ND_sim_onep_ex_con}}
\caption{Critical negative perturbation and extinction.}\label{2_var_nd_ex}
\end{figure}

Figures \ref{2_ND_sim_prey_onep_same_sol}, \ref{2_ND_sim_prey_onep_same_two_sol} and \ref{2_ND_sim_prey_onep_same_cons} show the reaction and recovery of the nonlinear interacting system from an external perturbation. One of the typical recovery of a system from a perturbed state is shown. In Figure \ref{2_ND_sim_prey_onep_same_sol}, an external perturbation starts at $t=700$ (Sp.1), and the coefficient $f_1$ equals to $-1260.0$ and $f_2$ equals to zero in this example. The black arrow is the starting point of perturbation, and the gray arrow is the end of perturbation in Figures \ref{2_ND_sim_prey_onep_same_sol} and \ref{2_ND_sim_prey_onep_same_cons}. The nonlinear coefficients are listed in Table 1 (Condition 1). The solutions $(x_1,x_2)$ are deformed by the perturbation (Figure \ref{2_ND_sim_prey_onep_same_sol} and \ref{2_ND_sim_prey_onep_same_two_sol}). However, the system does not disintegrate but finds a new stable phase-space close to the original phase-space and maintains a new conserved relation. The perturbation ends at $t=1200$ (Ep.1), and the system recovers the original state $(x_1,x_2)$. 

\begin{figure}[h]
\begin{center}
 \subfigure[2-variable ND solutions with a positive perturbation. The perturbation starts at $t=500$ and ends at $t=900$ represented as gray background. The amplitudes of $x_1$ and $x_2$ become larger than before.]{%
   \includegraphics[width=.44\columnwidth]{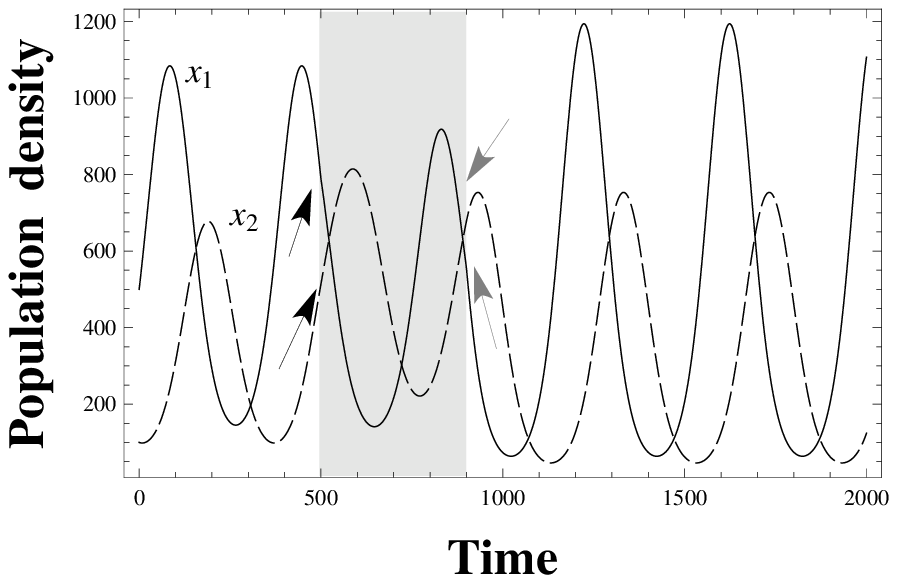}\label{2_ND_sim_onep_dvi_sol_ref}}\hspace{10mm}
 \subfigure[The Conserved quantity $\Psi$ with a positive perturbation. It recovers after perturbation but finds another equilibrium state.]{%
   \includegraphics[width=.45\columnwidth]{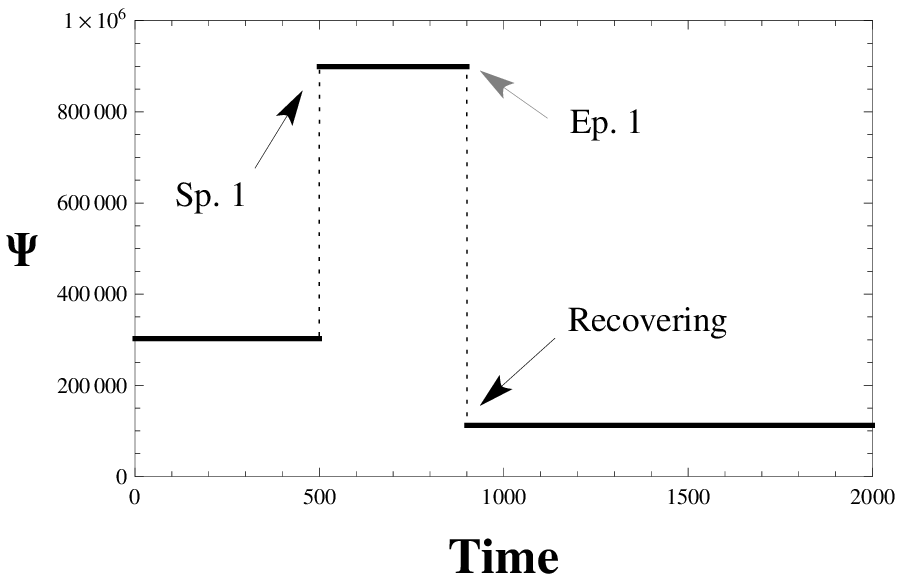}\label{2_ND_sim_onep_dvi_con_ref}}\\
\end{center}
 \subfigure[2-variable ND solutions with a critical perturbation. $x_1$ and $x_2$ converge to zero after a critical perturbation. ]{%
   \includegraphics[width=.43\columnwidth]{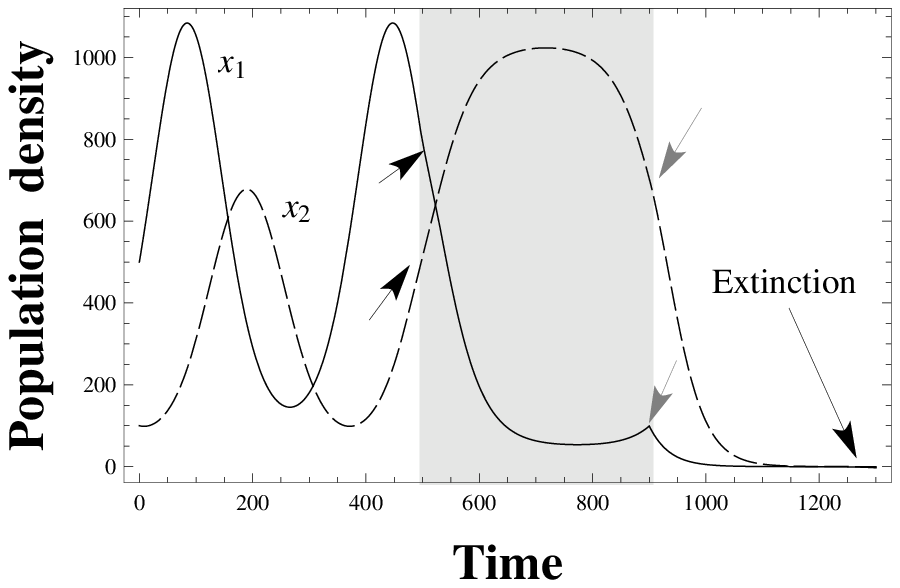}\label{2_ND_sim_onep_dvi_sol}}\hspace{13mm}
 \subfigure[Conservation law of 2-variable ND with a critical perturbation. The $\Psi$ converges to zero after a critical perturbation.]{%
   \includegraphics[width=.45\columnwidth]{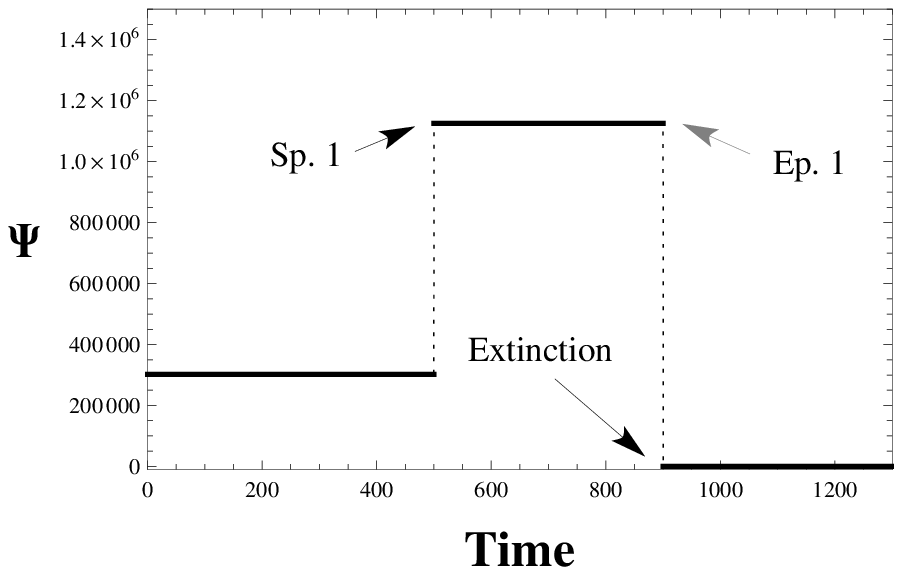}\label{2_ND_sim_onep_dvi_con}}
\caption{Critical perturbations and divergence of solutions.}\label{2_var_nd_div}
\end{figure}

The timing of negative perturbation which reduces the population number $x_1$ or $x_2$ produces different results. When a negative perturbation is exerted in the increasing phase of $x_1$ or $x_2$, the system will find a new conserved stable solution near the original solution, but when a negative strong perturbation is exerted before $x_1$ or $x_2$ gets to its minimum, the system may collapse: the system exhibits no solutions $(x_1, x_2)$, which would be interpreted as disintegration or extinction in biological systems.

The conserved nonlinear system naturally exhibits maxima and minima without external perturbations, and so we call these maxima and minima as endogenous maximum and minimum. It is needed to distinguish them from enhanced maxima and minima by external perturbations.

In Figure \ref{2_var_nd_ex}, the response of a strong negative perturbation to prey after the peak of endogenous maximum is shown. The values of coefficients are listed in Table 1 (Condition 1). The starting point of this perturbation is at $t=800$ and the end point of the perturbation is at $t=950$. The negative constant of perturbation is $f_1=-3175.3879$. The prey, $x_1$, rapidly declines with negative perturbation, and ($x_1$, $x_2$) converges to zero for $t \gtrsim 1000$.
These computer simulations may be compatible with known empirical results, for example, in pest control. A pest control is not so effective if it is performed in the season when harmful insects are in peak and active, because species are energetic enough to find a new stable life to live. It is effective when a pest control is performed in the season when harmful insects are not so active or in a declining state after endogenous maximum.

\begin{figure}[h]
 \subfigure[2-variable ND solutions with perturbations to avoid converging to zero after a critical perturbation. $x_1$ and $x_2$ come back to life after the second perturbations.]{%
   \includegraphics[width=.43\columnwidth]{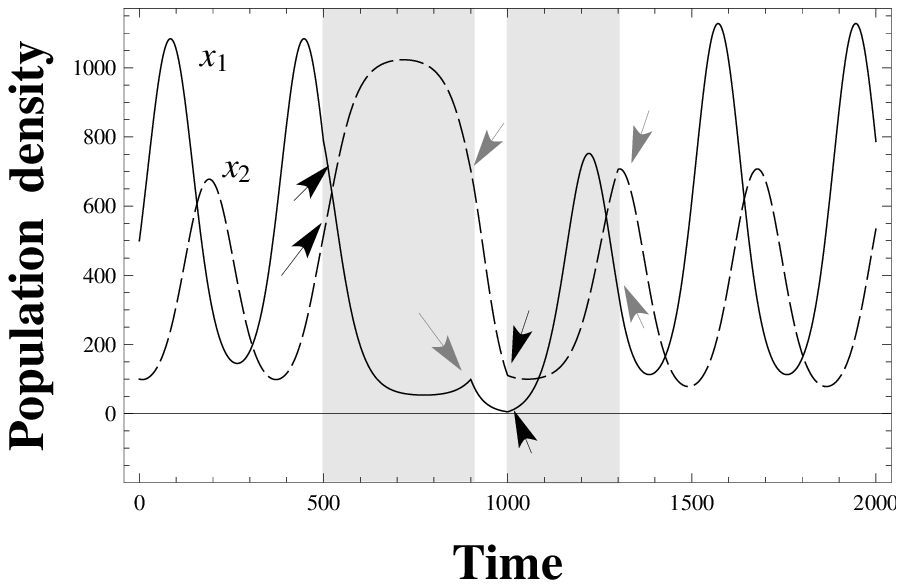}\label{2_ND_sim_onep_dvi_sol_save}}\hspace{10mm}
 \subfigure[Conservation law $\Psi$ with two perturbations in (a). $\Psi$ recovers from the perturbation after Sp.~2 - Ep.~2.]{%
   \includegraphics[width=.45\columnwidth]{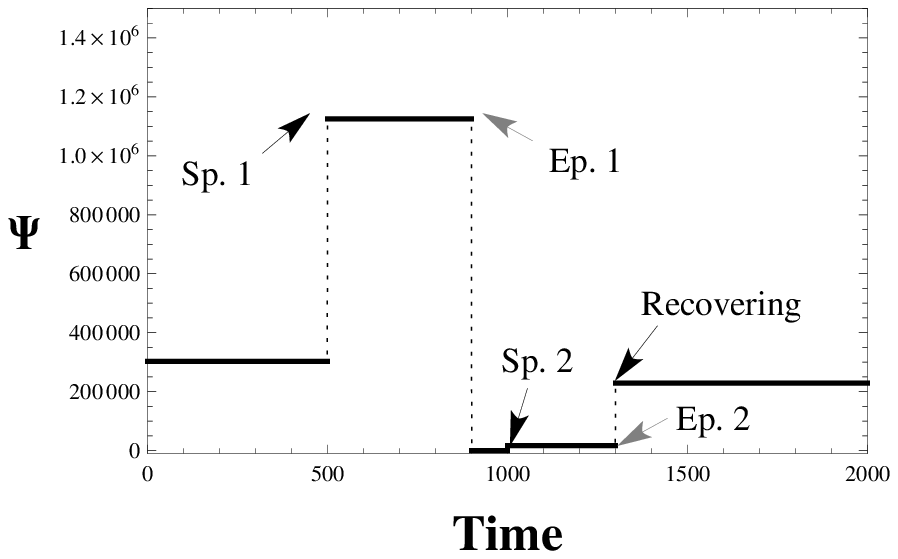}\label{2_ND_sim_onep_dvi_con_save}}
\caption{The critical behavior and restoration.}\label{2_var_nd_div_save}
\end{figure}

In the nonlinear interacting system, positive perturbations which will increase $x_1$ or $x_2$ do not always mean a positive effect on stability of the system. There is a limit to the value of a positive perturbation, because an increase of $x_1$ leads to a decrease of $x_2$ in a stable system $(x_1, x_2)$, which indicates that a system has internally allowed maximum and minimum populations.

Figures \ref{2_var_nd_div} shows the behaviors of $(x_1, x_2)$ at normal and critical values of positive perturbations, $c_1$, for $x_1$. The values of coefficients are listed in Table 1 (Condition 2). Figures \ref{2_ND_sim_onep_dvi_sol_ref} and \ref{2_ND_sim_onep_dvi_con_ref} show that the normal positive perturbation which increases interacting species will increase the peak of $(x_1, x_2)$ populations. However, at certain critical values of coupling constants, the prey-predator interaction cannot keep and support the rhythm of maxima and minima, and the system diverges. Figures \ref{2_ND_sim_onep_dvi_sol} and \ref{2_ND_sim_onep_dvi_con} show that the system cannot maintain a stable, interacting system when the positive perturbation surpasses the critical value ($c_1= 1599.924999$ in the current simulation). The unstable solutions branch out at $t \simeq 1100$ when the value of perturbation changes from $f_1=1160.0$ to $f_1=1599.924999$.

Hence, in a conserved stable system, species seem to strictly control each other by seeking a new stable solution so that they can survive together. The competitive interacting system such as the conserved prey-predator relations may be considered to be a cooperative system for species to survive. It should be noted that if a dynamical prey-predator system is active, the rhythms of maxima and minima are clearly repeated, which is known in real prey-predator systems. However, if an external perturbation (exogenous interaction) exceeds a certain critical value of the competitive system, the rhythms of maxima and minima will disappear first and then after a time, the system will diverge (disintegrate). Therefore, the rhythm of wild-life indicates that the dynamical interactions between species are active and stable. When the rhythm of change disappear or doesn't come back, it may indicate that related species are in danger of extinction. The rhythm is important to examine if the wild life is normal and active, or harmed by human activities and external perturbations.

On the other hand, by adding another perturbation, we can show that it is possible to save species from extinction. Figure \ref{2_ND_sim_onep_dvi_sol_save} is a result of a positive perturbation to save species $(x_1, x_2)$ in a danger of extinction in Figure \ref{2_ND_sim_onep_ex_sol}. We exerted a positive perturbation after Sp.~1 - Ep.~1 in Figure \ref{2_var_nd_div_save}. The positive perturbations start at $t=1000$ (Sp.~2) and end at $t=1300$ (Ep.~2), the strengths of $c_1$ and $c_2$ are $f_1 = 200$, $f_2 = -1000$. Figure \ref{2_ND_sim_onep_dvi_sol_save} shows that species are in danger of extinction, however, if positive external perturbations are properly inserted, the system will come back to life again.

\begin{figure}[h]
\begin{center}
 \subfigure[2-variable ND solutions with three external perturbations. The rhythm of $x_1$ and $x_2$ recovers from several perturbations. Gray backgrounds represent periods of perturbations.]{%
   \includegraphics[width=.45\columnwidth]{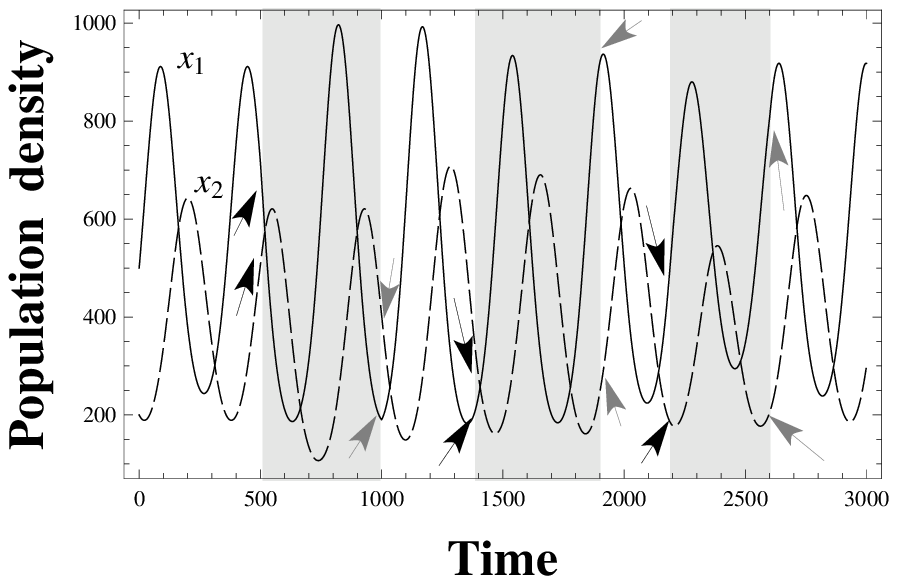}\label{2_ND_sim_prey_threep_same_sol}}\\
\end{center}
 \subfigure[Phase-space transitions of $x_1$ and $x_2$. Dashed lines represent solutions, $x_1$ and $x_2$, during perturbations. Solid line represent solutions without perturbations.]{%
   \includegraphics[width=.43\columnwidth]{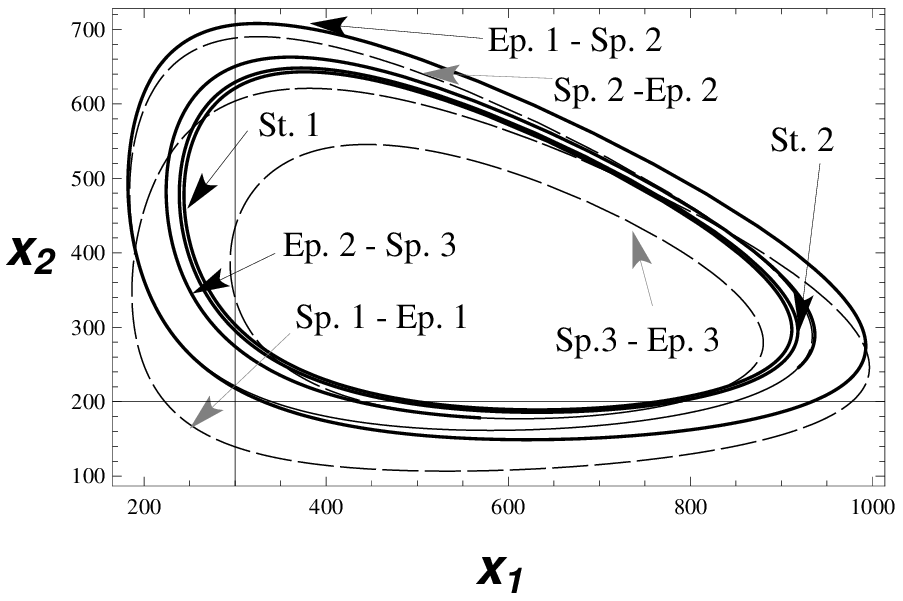}\label{2_ND_sim_prey_threep_same_two_sol}}\hspace{10mm}
 \subfigure[Conservation law $\Psi$ with three perturbations. It recovers from three perturbations. $\Psi \simeq 60000$ in the St.~1 and St.~2.]{%
   \includegraphics[width=.45\columnwidth]{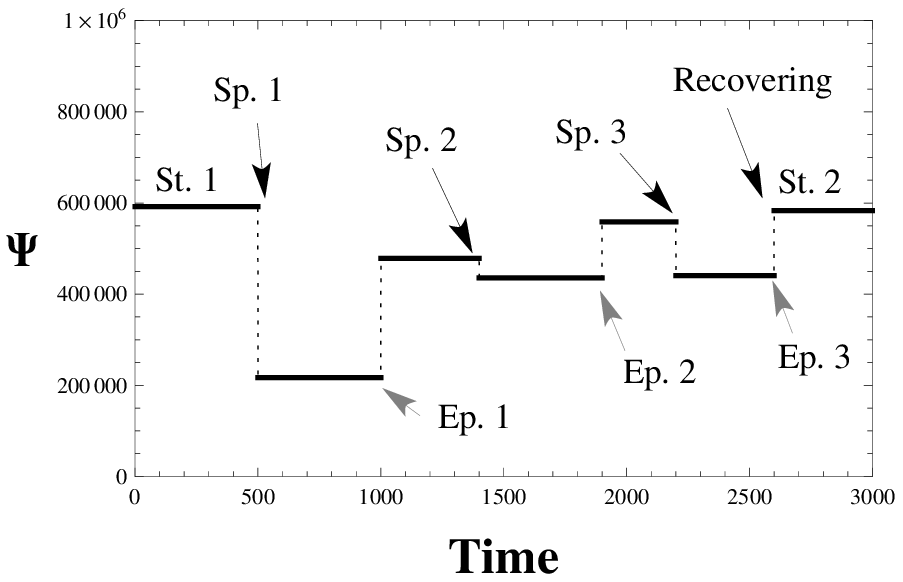}\label{2_ND_sim_prey_threep_same_cons}}
\caption{Several external perturbations and recoveries.}\label{2_var_3_nd}
\end{figure}

\section{Conservation law and population cycles}\label{Conservation law with external perturbation}
\subsection{The food-web of Microbes in Okanagan Lake}\label{The food-web of Microbes in Okanagan Lake}

One of interesting data of the ecological interactions is the interaction described in \lq {\it Mysis} and the other zooplanktons in the Okanagan Lake\rq \cite{schindler2012mysis}. Although the food-web in Okanagan Lake is not clarified definitely, mysis introduction to lakes is known as an effective method to enhance ecological interactions and its strengths among microbes and other creatures so as to increase fisheries productions.

The time-series of dominant crustacean zooplankton densities in Okanagan lake has been measured monthly and suggested that mysis and zooplankton populations are synchronous and characterized by the cycle of the peak and bottom population densities.  The cycles of population densities are primarily due to cycles of season and climate and then to mutual interaction of microbes. The analysis of microbes suggests that the density-dependent and delayed population regulation of microbes is evident. In addition to the seasonal factors, the regular cycles and the delayed peak and bottom populations densities of microbes are the results of strong nonlinear interactions of species. We numerically examined changes of population densities of microbes by employing the 2-variable conserved ND model.

The current conserved nonlinear model shows that the interacting species designate a {\it standard rhythm} of the peak and bottom population densities. There are some fluctuations at the peak and bottom densities, but they show the stable dynamic life as demonstrated in Figure \ref{2_var_3_nd} (a) $\sim$ (c). Although, normal peak and bottom densities can be readily explained by adjusting coupling strength of model's internal interactions, a sudden change of maxima which is often encountered in a biological data cannot be easily simulated by only adjusting internal coupling constants in the 2-variable nonlinear interacting model.

In Figure \ref{2_var_3_nd}, several perturbations are exerted on the interacting 2-variable system. The first external perturbation starts at $t=500$ (Sp.~1) and ends at $t=1000$ (Ep.~1). The strength of perturbations in Sp.~1-Ep.~1 are $f_1 = -800$, $f_2=-100$. The second external perturbation starts at $t=1400$ (Sp.~2) and ends at $t=1900$ (Ep.~2). The  strengths of perturbations in Sp.~2-Ep.~2 are $f_1 = -50$, $f_2=-120$. The third external perturbation starts at $t=2200$ (Sp.~3) and ends at $t=2600$ (Ep.~3). The strength of perturbations in Sp.~3-Ep.~3 is set as $f_1=-500$, $f_2=-50$. The lines $(x_1,x_2)$ may represent for instance, the prey-predator interactions, species of food-chain, and species interacting with its environmental factors (temperature or some environmental hormones). Black arrows are starting point of perturbations, and
gray arrows are the end of perturbations; parameters are listed in Table 1 (Condition 1). The time period is within $t=4000$, initial values are $x_1 = 500$, $x_2 = 300$.

The significant properties of the stable nonlinear conserved system are that if external perturbations are not large enough to disintegrate the system, the system will find a stable conserved solution near the original system and continue a stable cycle (maxima and minima). It is clearly seen from $(x_1,x_2)$-phase space solutions in Figure \ref{2_ND_sim_prey_threep_same_two_sol}. The system recovers from several external perturbations.

The numerical analysis can be applied to examine the change of population densities of microbes. For example, the time-series data of dominant crustacean zooplankton densities in the Figure 2 of the paper \lq {\it Mysis} in the Okanagan Lake food-web \rq, show that the sudden maxima of dominant zooplankton densities are seen in the period '99 $\sim$ '02. The sudden increase of the peak is readily adjusted when an external perturbation is assumed in the simulation, however, it is not reproduced by adjusting internal coupling constants in the 2-variable nonlinear model. Hence, it is concluded in the 2-variable model that there would have been certain positive external perturbation to the system of microbes in Okangan Lake during '98 $\sim$ '01 considering a time-delay of external perturbations.

It is interesting to check what kind of external or internal perturbations is affecting the peak of population density during the period  '98 $\sim$ '01. If there are no explicit changes in external or internal factors during the period, a sudden increase of the peak could be a result of more complex internal interactions. For example, the rhythm of the peak and bottom population densities should be explained by 4-variable or 6-variable nonlinear interactions of microbes.
The unusual rhythm indicates how exogenous (environmental) and endogenous (internal interactions) variables are affecting the dynamics of each component and environmental nature related to the species. The analysis of nonlinear model suggests that the sudden peak and bottom densities have important information on the dynamics of the system of species and environment. Hence, it is important to understand the {\it standard rhythm} of the peak and bottom population densities in order to distinguish them from unusual maxima and minima.

One should be careful that a positive perturbation on one of interacting species not only enhances the peak of maxima but also decreases minima in the rhythm of species. It is often true that the effect of enhancement is usually emphasized without taking care after negative effects. Hence, the enhancement of the number of population of a specific species may be harmful to other species in the food-web and consequently it endangers itself. Our analyses in Figure \ref{2_var_nd_div} and \ref{2_var_nd_div_save} show that if we carefully control the increase or decrease of the population of certain species after introduction of a positive effect, we can keep normal and stable dynamics of species suitable for the environment. For this purpose, it is essential to explicitly understand the {\it standard rhythm} from real observed data. 

\subsection{Population regulation in Canadian lynx and snowshoe hare}\label{Population regulation in Canadian lynx and snowshoe hare}

It is difficult to identify population regulation mechanisms about prey-predator patterns of large mammals because the large mammal's life span is relatively long compared with microbes. The prey-predator cycle such as wolves and caribous takes some decades of years to observe, their interacting relation and behaviors have been recently revealed with modern technology (GPS-colored animals) \cite{stenseth1998patterns}.
However, the food-web configuration between snowshoe hare and Canadian lynx is well-known prey-predator type phenomena, and a ten-year cycle of Canadian lynx was examined from the data of Canada lynx fur-trades return of the Northern Department of the Hudson's Bay Company (the data are from C. Elton and M. Nicholson \cite{elton1942ten}). 

\begin{figure}[h]
\ \subfigure[The 2-variable ND simulation of Canadian lynx population. The solid line represents Canadian lynx population \cite{elton1942ten}, and the dashed line represents a theoretical solution of 2-variable ND with several perturbations.]{%
   \includegraphics[width=.43\columnwidth]{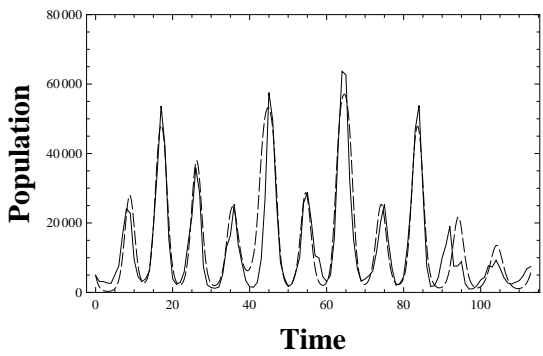}\label{2_ND_sim_lynx_data}}\hspace{10mm}
 \subfigure[The estimated population of Canadian lynx and snowshoe hare. The dashed line represents Canadian lynx population simulated by 2-variable ND model with perturbations, and the solid line represents approximate population of snowshoe hare.]{%
   \includegraphics[width=.45\columnwidth]{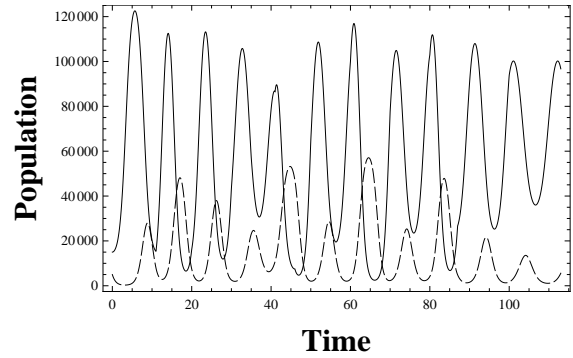}\label{2_ND_sim_lynx_data_x1x2}}
\begin{center}
 \subfigure[Transition of conservation law $\Psi$ with respect to time. Several perturbations are introduced.]{%
   \includegraphics[width=.45\columnwidth]{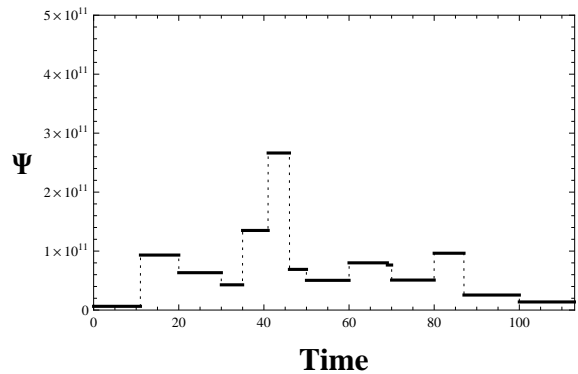}\label{2_ND_sim_lynx_data_cons}}
\end{center}
\caption{Simulation of Canadian lynx and snowshoe hare.}\label{2_var_sim1}
\end{figure}

\begin{table}[b]
\centerline{
Table 2: The list of nonlinear coefficients of simulation in Figure \ref{2_var_sim1}.}\vspace{0.2cm}
\begin{center}
  \begin{tabular}{c |c| c| c | c|  c| c| c} \hline
 $\alpha_1$  & $\alpha_2$ & $\alpha_3$ & $\alpha_4$ & $\alpha_5$ & $\alpha_6$ & $\alpha_7$ & $\alpha_8$ \ \\  \hline \hline1.0 & 635.0  & -0.35 & 700.5 & 300.5 & -0.35 & -0.0068 & -0.016 \\ \hline
      \end{tabular}
\end{center}
\end{table}

\begin{table}[h]
Table 3: The list of external perturbations in Figure \ref{2_var_sim1}.
 The periods of positive and negative perturbations to numerically simulate Canadian lynx population. Note that the values of $f_1$ have the meaning of velocity (number/time).\vspace{0.1cm}
\begin{center}
  \begin{tabular}{c | c | c} \hline
Perturbation & Time & Strength of $f_1$ \\ \hline \hline
Sp.~1 - Ep.~1 & 0 $\leq$ t $\leq$ 11 & -11000000 \\ \hline
Sp.~2 - Ep.~2 & 20 $\leq$ t $\leq$ 30 & -4000000 \\ \hline  
Sp.~3 - Ep.~3 & 30 $\leq$ t $\leq$ 35 & -12000000 \\ \hline
Sp.~4 - Ep.~4 & 35 $\leq$ t $\leq$ 41 &  -8000000 \\ \hline
Sp.~5 - Ep.~5 & 41 $\leq$ t $\leq$ 46 &  4200000 \\ \hline
Sp.~6 - Ep.~6 & 50 $\leq$ t $\leq$ 60 & -9000000 \\ \hline
Sp.~7 - Ep.~7 & 60 $\leq$ t $\leq$ 69 & 1140000 \\ \hline
Sp.~8 - Ep.~8 & 70 $\leq$ t $\leq$ 80 & -11000000 \\ \hline
Sp.~9 - Ep.~9 & 87 $\leq$ t $\leq$ 100 & -14500000 \\ \hline
Sp.~{10} - Ep.~{10} & 100 $\leq$ t $\leq$ 113 & 21500000 \\ \hline
      \end{tabular}
\end{center}\label{table1}
\end{table}

The Canadian lynx and snowshoe hare have a synchronous ten-year cycle in population numbers \cite{stenseth1998patterns,stenseth1997population}. 
The fundamental mechanisms for these cycles are maintained by the important factors such as nutrient, predation and social interactions \cite{krebs2001drives}.
In addition to the important factors, the nonlinear model with conservation law suggests that species of a system consequently find a strategy or a mechanism to survive for long-time periods. In other words, the cycle of population density is a manifestation of the strategy or mechanism to survive, which is suggested by stability of phase-space solutions determined by conservation law of a system.

The nonlinear interactions with conservation law show a {\it standard rhythm} and stability from external perturbations as shown in Figure \ref{2_var_3_nd}. The feeding and nutrient experiments in \cite{krebs2001drives} are considered as external perturbations to the system. As shown in Figure \ref{2_var_3_nd}, the perturbations cause certain effects on the system, but the system will find a rhythm to maintain the dynamics of species, which is not so different from the original {\it standard rhythm}. Our numerical results agree with conclusions derived from feeding experiments and nutrient-addition experiments. Therefore, we propose that the properties of the system which has a conservation law should be a key to understand the unanswered question: why do these cycles exist?.

The results of computer simulations show that the timing of perturbation leads to different results. This is also confirmed by the feeding experiment of snowshoe hare: `` ... during the peak of the cycle in 1989 and 1990 had no impact on reproductive output ... however, during the decline phase in 1991 and 1992, the predator exposure plus food treatment caused a dramatic increase in reproductive output ..." \cite{krebs2001drives}.
This fact can be examined in our model calculations. The perturbation in the peak phase does not cause large effects on {\it standard rhythm}, but negative and positive perturbations during a decreasing or increasing phase induce dramatic effects.

The cycle of {\it standard rhythm} for Canadian lynx and snowshoe hare indicates that the stable dynamical system of lynx and hare functions normally and environmental nature is conserved in reasonable conditions. However, as we have shown in Figure \ref{2_ND_sim_onep_dvi_sol} and \ref{2_ND_sim_onep_dvi_con}, if a strong negative perturbation is applied persistently for a long period, the system would fall into a danger of extinction.
The important results of our simulation tell that before a system gets in danger of extinction, the {\it standard rhythm} of the system will tend to become ambiguous or disappear.
Hence, if we carefully observe the {\it standard rhythm} of a specific system of species, we could help the dynamical system save and preserve related natural environment.

In Figure \ref{2_var_sim1}, we simulated the Canadian lynx data of the Hudson's Bay Company from 1821 to 1910, which is approximately thought as the lynx-population density. The interpolated Elton\rq s data was downloaded from \url{http://www.atomosyd.net/spip.php?action=dw2_out&id=42}. The solid-line in Figure \ref{2_ND_sim_lynx_data}, is lynx-population data and the dashed-line is the results of our numerical simulation using 2-variable nonlinear interactions between lynx and snowshoe hare (Figure \ref{2_ND_sim_lynx_data_x1x2}). The conserved binary-coupled model tells that there should have been some external perturbations, although we cannot make sure at the present what kinds of external perturbations were exerted. The actual population density of snowshoe hare is not known, and so we assumed a reasonable population density and several external perturbations for numerical simulations in order to fit the lynx population data (see, Table 2 and Table 3).

The snowshoe hare gets several positive and negative perturbations, but the overall rhythms of lynx and hare are not altered. As suggested by in Figure \ref{2_ND_sim_prey_threep_same_two_sol}, the phase-space of lynx and snowshoe hare is stable against several external perturbations. This is also compatible with the empirical fact that the ten-year cycle in snowshoe hare is resilient to a variety of natural disturbances from forest fires to short-term climatic fluctuations. However, as shown in our model calculation in Figure \ref{2_ND_sim_onep_dvi_sol}, a long-term (more than ten years) negative perturbations and a vast environmental change that humans could cause would definitely endanger the standard rhythm of snowshoe hare, lynx and related species. 
\section{Conclusions}\label{Conclusion and Discussion}
In this paper, we examined characteristic properties of several ecological systems based on conserved nonlinear interactions which include generalized Lotka-Volterra type prey-predator, competitive interactions. In Section \ref{2n-ND system with perturbations}, we extended our $2n$-variable ND model by including external perturbations in order to apply the model to more realistic biological phenomena and to study responses of a biological system to external perturbations.

We simulated external positive and negative perturbations by employing piecewise constant terms in our nonlinear equations. As it is discussed in the analysis, the results of simplified perturbations agreed with the experiments and empirical data reasonably well. The numerical simulations showed the existence of the {\it standard rhythm} which is characteristic to a nonlinear conserved system. It is essential to understand {\it standard rhythm} by observing and taking data of a system so that we can distinguish unusual maxima and minima from {\it standard rhythm}. This gives a possibility to examine signatures that distinguish internal effects from external ones.

The ten-year cycle of lynx and hare is a very interesting biological phenomena. Though a cycle of a biological system should be a phenomenon composed of complex and multi-biological interactions, the 2-variable BCF analysis has revealed the interesting results on properties of the biological phenomena. The ten-year cycle of lynx and hare is stable and resilient to external perturbations, which is reproduced in our model calculations. The system with conservation law shows stable cycles and recovering phenomena, which are displayed numerically in phase-space solutions. The stability and conservation law are constructed at least by binary-coupled species in biological and ecological systems, and they are maintained in a more complicated multi-coupled system, as we proved in a general form \cite{uechi2012conservation}.

The coupling constants of interacting species expressed in nonlinear differential equations are considered to have been determined in a long time by complicated environmental and internal factors of a specific system, such as the landforms, seasons, climate and temperature. Once members and structures of dynamical systems were constructed, appropriate dynamical systems would be maintained for long-time periods with internal factors such as nutrient, predation and social interactions. The predation and social interactions are expressed as complicated nonlinear relations in mathematical terms. This may be explained by the fact that members of a system have a well-conserved rhythm respectively and these rhythms also have a well-determined slight delay to each other, which indicates that certain nonlinear interactions among members exist.

The important factors (nutrient, predation and social interactions) are needed for all species to survive in nature, but they easily change by natural conditions. In addition, an unusual increase of population numbers of a species would endanger the survival of a species itself as well as other species (see the numerical simulations in Figure \ref{2_var_nd_div}). The important property of the nonlinear model with conservation law is that the binary-coupled system can have the persistent stability and recovering strength to external perturbations. As a predator needs a prey for its food, a prey needs a predator for the conservation of their own species. The conservation law and rhythm of species are considered to be constructed by species and natural conditions in a system for a long time, and hence, the cycle (rhythm) of species would be interpreted as  a manifestation of the survival of the fittest to the balance a biological system.

We conclude that stability and conservation law are constructed by species in mutual dependency or cooperation to survive for long-time periods in severe nature. The standard rhythm should be regarded as the result of strategy for species to live in nature. Whatever roles they have to play, the species that can fit and balance with other creatures can survive in nature. A strong predator cannot even survive if it ignores the law of the standard rhythm and conservation law of a system constructed by other members and the environment. We hope that this study will help understand both activities of animals and humans in natural life.
\newpage
\nocite{*}
\bibliography{conserved_recovery}

\end{document}